# A Study of Primordial Very Massive Star Evolution

Guglielmo Volpato[1], Paola Marigo[1], Guglielmo Costa[1,2,3], Alessandro Bressan[4], Michele Trabucchi[1,5], and Léo Girardi[3]

[1] Dipartimento di Fisica e Astronomia Galileo Galilei, Università degli studi di Padova, Vicolo dell'Osservatorio 3, I-35122 Padova, Italy  
guglielmo.volpato@phd.unipd.it  
[2] INFN-Padova, via Marzolo 8, I-35131 Padova, Italy  
[3] Osservatorio Astronomico di Padova-INAF, Vicolo dell'Osservatorio 5, I-35122 Padova, Italy  
[4] SISSA, via Bonomea 365, I-34136 Trieste, Italy  
[5] Department of Astronomy, University of Geneva, Ch. des Maillettes 51, 1920 Versoix, Switzerland

## Abstract

We present new evolutionary models of primordial very massive stars with initial masses ranging from 100 to 1000 $M_\odot$ that extend from the main sequence to the onset of dynamical instability caused by the creation of electron–positron pairs during core C, Ne, or O burning, depending on the star's mass and metallicity. Mass loss accounts for radiation-driven winds, as well as pulsation-driven mass loss on the main sequence and during the red supergiant phase. After examining the evolutionary properties, we focus on the final outcome of the models and associated compact remnants. Stars that avoid the pair instability supernova channel should produce black holes with masses ranging from $\approx 40$ to $\approx 1000\, M_\odot$. In particular, stars with initial masses of about 100 $M_\odot$ could leave black holes of $\simeq 85$–90 $M_\odot$, values consistent with the estimated primary black hole mass of the GW190521 merger event. Overall, these results may contribute to explaining future data from next-generation gravitational-wave detectors, such as the Einstein Telescope and Cosmic Explorer, which will have access to an as-yet-unexplored black hole mass range of $\approx 10^2$–$10^4\, M_\odot$ in the early universe.

*Unified Astronomy Thesaurus concepts:* Stellar evolution (1599); Stellar remnants (1627); Stellar winds (1636); Population III stars (1285)

## 1. Introduction

For the first generations of stars (Population III), the efficiency of the cooling processes that regulate star formation is considerably diminished due to the absence or severe deficit of metals. In the early universe, magnetic fields and turbulence might also be less significant (Abel et al. 2002). As a consequence, for primordial stars in their unique conditions, the minimum mass for fragmentation (the local Jeans mass) may have been as high as $\simeq 1000\, M_\odot$ (e.g., Larson 1998; Hosokawa et al. 2011; Hirano et al. 2014; Stacy et al. 2016). From numerical simulations of star formation, we expect that such stars form at redshift $z \simeq 20$ and have initial mass functions that either peak at $\simeq 100\, M_\odot$ (Bromm et al. 1999; Abel et al. 2002) or present a bimodal distribution with a second peak at a few $M_\odot$ (Nakamura & Umemura 2001). Other studies, in contrast, claim that the characteristic mass of the Population III initial mass function could be significantly lower than the canonical 100 $M_\odot$ (Clark et al. 2011).

Extremely metal-poor or zero-metallicity very massive stars, with initial masses in the range $100 \lesssim M_i/M_\odot \lesssim 1000$, have a broad astrophysical impact. Understanding how these Population III stars evolve and die has implications for several key questions, including the observable characteristics of integrated stellar populations in low-metallicity galaxies; the nature of energetic transients, such as pair instability supernovae (PISNe), superluminous supernovae, kilonovae, and gamma-ray bursts (Kozyreva et al. 2017); the source of extreme ionizing UV radiation fields at high redshift (Dijkstra & Wyithe 2007); the agents of the earliest and fastest chemical enrichment of their host galaxies (Kozyreva et al. 2014; Goswami et al. 2021, 2022); the rates of gravitational-wave emission from merging black holes (BHs; Abbott et al. 2016; Spera et al. 2019); and the formation of primordial stellar BHs that could provide the seeds for the assembly of supermassive BHs of mass $\simeq 10^6$–$10^9\, M_\odot$ at redshift $z > 6$ via runaway stellar collisions in dense clusters (Belkus et al. 2007; Sakurai et al. 2017; Onoue et al. 2019; Nakauchi et al. 2020).

Very massive stars ($100 \lesssim M_i/M_\odot \lesssim 300$) may undergo electron–positron pair instabilities (PIs) before and during core oxygen burning, with a final outcome determined primarily by the mass of the helium core, $M_{He}$, eventually leading to a successful or failed core-collapse supernova (CCSN) or thermonuclear explosion (Heger & Woosley 2002; Kozyreva et al. 2017; Woosley 2017; Leung et al. 2019).

Stars with final helium core masses in the approximate range $34$–$45 \lesssim M_{He}/M_\odot \lesssim 64$ are predicted to join the domain of pulsation pair instability supernovae (PPISNe). During these unstable stages, several strong pulses may eject a significant fraction of the star's residual envelope and possibly a small fraction of the core mass before dying with a successful or failed CCSN (Woosley et al. 2002; Chen et al. 2014; Yoshida et al. 2016; Woosley 2017; Farmer et al. 2019; Woosley & Heger 2021; Farag et al. 2022). Stars with larger helium core masses, $64 \lesssim M_{He}/M_\odot \lesssim 135$, are predicted to die as PISNe. Near the ignition of core oxygen burning, such stars experience a single strong pulse followed by a thermonuclear explosion that unbinds the whole star, leaving no remnant (Fowler & Hoyle 1964; Barkat et al. 1967; Rakavy & Shaviv 1967; Fraley 1968; Heger & Woosley 2002; Heger et al. 2003; Takahashi 2018; Takahashi et al. 2018; Woosley & Heger 2021;

Farag et al. 2022). In the past, PISNe have been traditionally associated with the first, extremely metal-poor stellar populations (Ober et al. 1983; Bond et al. 1984; Glatzel et al. 1985; Heger & Woosley 2002), though recent stellar models suggest that PISNe could happen for stars with initial metallicity up to $Z \simeq Z_\odot/3$ (Langer et al. 2007; Yusof et al. 2013; Kozyreva et al. 2014). The Tarantula Spectroscopic Survey (Schneider et al. 2018; Crowther 2019), which indicates that the current initial mass function is well populated up to 200 $M_\odot$ in the Large Magellanic Cloud, lends support to the existence of very massive stars at these metallicities.

Stars massive enough to form a helium core with $M_{\rm He} \gtrsim 135\,M_\odot$ are predicted to undergo direct collapse to a BH (DBH). During the final stages, photodisintegration processes absorb the energy of the propagating shock, preventing the envelope from becoming unbound through mass ejection (Bond et al. 1984; Farmer et al. 2020). In these circumstances, only wind ejecta are produced (Fryer & Kalogera 2001; Heger & Woosley 2002; Nomoto et al. 2013). In this framework, we expect stars with $M_{\rm i} \gtrsim 200\text{--}300\,M_\odot$ to avoid the thermonuclear explosion at very low metallicity (e.g., Goswami et al. 2021).

The details of this evolutionary picture, particularly the ranges of initial masses of stars that follow the same channel and achieve a similar final outcome, are affected by factors such as metallicity and the efficiency of stellar winds, among others (e.g., Vink et al. 2021). Indeed, mass loss is a critical process in the evolution of massive and very massive stars, though some aspects are still not completely understood. Mass loss contributes significantly to the chemical enrichment of the interstellar medium, can affect star formation by injecting momentum and kinetic energy into molecular clouds, and may have a decisive impact on the outcome of core collapse.

The winds of massive and very massive stars can be triggered and maintained by a variety of physical processes (Renzo et al. 2017). In hot and luminous stars, the radiation field transfers momentum to the outflowing plasma via scattering in resonant spectral lines (e.g., Vink et al. 2001; Puls et al. 2008). Continuous absorption and scattering from dust grains act in the extended circumstellar envelopes of luminous red supergiants (RSGs), in which the interplay between pulsation and near-surface turbulent convection can also be important for mass loss (e.g., Bennett 2010; Höfner & Olofsson 2018; Kee et al. 2021). Luminous blue variables also involve pulsational mass loss alongside eruptive phenomena (Baraffe et al. 2001; Puls et al. 2008; Smith & Arnett 2014; Nakauchi et al. 2020), whereas mass loss is associated with Roche-lobe overflow and common-envelope evolutionary phases in interacting binary systems (e.g., Woosley et al. 1995; Wellstein & Langer 1999; Smith & Tombleson 2015; Shara et al. 2017).

The theory of line-driven winds applied to hot and luminous stars predicts a positive correlation between the mass-loss rate and the metal content (e.g., Vink et al. 2011), which implies that stellar winds in extremely metal-poor conditions should be quite weak (see, for instance, the $Z = 0$ models of Marigo et al. 2003). We know that primordial main-sequence (MS) stars are pulsationally unstable above a critical mass of $\simeq 100\,M_\odot$ due to the destabilizing effect of nuclear reactions in their cores (Baraffe et al. 2001; $\epsilon$-mechanism). The instability could reoccur during the core helium-burning (cHeB) phase, excited by the $\kappa$-mechanism operating in the hydrogen ionization zone.

Nonlinear calculations show that such an instability causes mass loss rather than catastrophic disruption. Nakauchi et al. (2020) recently performed a new stability analysis of primordial very massive stars during the MS and cHeB stages and provided analytic prescriptions for calculating the associated mass-loss rates.

Adopting the Nakauchi et al. (2020) results in combination with a well-tested scheme to account for radiation-driven winds, in this study, we use the PARSEC code to follow the evolution of very massive stars at zero and extremely low metallicity until the onset of dynamical instability caused by the creation of electron–positron pairs. We investigate the main evolutionary properties of these stars and predict their final outcome, which could be a massive BH or total incineration via a thermonuclear explosion.

The paper is organized as follows. In Section 2, we briefly describe the PARSEC code and its major ingredients. In Section 3, we present the stellar evolution models computed with mass-loss recipes that account for both radiation- and pulsation-driven mass loss. We provide an overview of evolutionary properties, with an emphasis on core evolution, dredge-up (DUP) episodes, internal structure, surface elemental abundances, chemical ejecta, final evolution outcomes, and associated compact remnants. Finally, Section 4 closes the paper with some concluding remarks and future perspectives.

## 2. Stellar Evolutionary Calculations

Stellar evolutionary models are computed with the PARSEC code version 2.0, as described in Bressan et al. (2012), Costa et al. (2019, 2021), and references therein. The main input physics and other ingredients used in the evolutionary calculations are summarized below.

The FREEEOS code developed by A. W. Irwin[6] is used to calculate the equation of state. Using the procedure described in Timmes & Arnett (1999), we include the effect of pair creation in the equation of state. Radiative opacities are taken from the OPAL project (Iglesias & Rogers 1996) for high temperatures, and the ÆSOPUS code (Marigo & Aringer 2009) is used for low temperatures. Conductive opacities are included according to Itoh et al. (2008). Nuclear reaction rates—p–p chains, CNO tri-cycle, Ne–Na and Mg–Al chains, and the most important $\alpha$-capture reactions, including $(\alpha, n)$ processes—together with the corresponding $Q$-values are taken from the JINA reaclib database (Cyburt et al. 2010). We use the fitting formulae by Haft et al. (1994) for plasma neutrinos, and we follow Munakata et al. (1985) and Itoh & Kohyama (1983) to account for energy losses by electron neutrinos. To describe mixing, we use the mixing-length theory (Böhm-Vitense 1958) with a fixed mixing-length parameter, $\alpha_{\rm ML} = 1.74$, calibrated on the present-day Sun's properties (Bressan et al. 2012). To test stability against convection, we use the Schwarzschild criterion. We apply core overshooting[7] described by the parameter $\lambda_{\rm ov} = 0.4$. For convective envelopes, we use an undershooting distance of $\Lambda_{\rm env} = 0.7\,H_P$ below the deepest unstable layer.

We consider two choices of the initial chemical composition defined by $(Z = 0, Y = 0.24850)$ and $(Z = 0.0002, Y = 0.24885)$, where $Z$ and $Y$ denote the initial abundances of metals and helium, respectively, and seven values of the initial mass, $M_i = 100, 150, 200, 300, 500, 750$, and $1000 M_\odot$. The $Y$-values are obtained using a primordial helium abundance of $Y_p = 0.2485$ (Komatsu et al. 2011) and a helium-to-metals enrichment ratio $\Delta Y/\Delta Z = 1.78$ based on the current PARSEC solar calibration (Bressan et al. 2012). For $Z = 0.0002$, the initial metal abundance distribution scales with the solar composition of Caffau et al. (2011).

---

[6] http://freeeos.sourceforge.net
[7] Here $\lambda_{\rm ov}$ is the mean free path of the convective element across the border of the unstable core in units of pressure scale height, $H_P$.

For each combination ($M_i$, $Z$), we apply three recipes to describe the mass-loss rate, $\dot{M}$, by stellar winds, namely,

1. $\dot{M} = \dot{M}_{rdw}$: radiation-driven winds as implemented in PARSEC (Section 2.1);
2. $\dot{M} = \dot{M}_{pdw}$: pulsation-driven winds according to the formulation of Nakauchi et al. (2020; see also Section 2.2); and
3. $\dot{M} = \dot{M}_{max}$: the highest rate between the above two cases, $\dot{M}_{max} = \max(\dot{M}_{rdw}, \dot{M}_{pdw})$.

In total, we produced six sets of stellar models, each defined by the initial metallicity, $Z$, and mass-loss prescription, $\dot{M}$. Since models with $\dot{M}_{pdw}$ were mainly intended to explore the sensitivity of pulsation-driven mass loss to stellar mass and effective temperature, in the analysis that follows, we will focus on the four sets computed with $\dot{M}_{rdw}$ and $\dot{M}_{max}$.

### 2.1. Radiation-driven Winds

Here we briefly review the standard mass-loss prescription adopted in PARSEC to treat mass loss from single massive stars in both the hot and cool regions of the H-R diagram (HRD). Details can be found in Chen et al. (2015), and some recent revision is described in Costa et al. (2021). For simplicity, we refer to it as the "radiation-driven winds" recipe, though we acknowledge that in RSGs, in addition to radiation on dust grains, other mechanisms may be at work, such as turbulence and pulsation.

In short, we rely on four main formulations. For hot stars with $T_{eff} > 10{,}000$ K, we adopt the formalism of Vink et al. (2000, 2001). We take into account the enhancement of the mass-loss rate when the Eddington factor, $\Gamma_e$, approaches unity (Gräfener & Hamann 2008; Vink et al. 2011). This parameter is commonly defined as

$$\Gamma_e = \frac{L \, \kappa_{es}}{4\pi \, G \, M}, \quad (1)$$

where $M$ and $L$ denote the current mass and luminosity, $\kappa_{es}$ is the opacity due to electron scattering, and $G$ is the gravitational constant. We use the same metallicity scaling relation as in Chen et al. (2015), which reads

$$\dot{M} \propto (Z/0.02)^\alpha, \quad (2)$$

with $\alpha$ given by

$$\alpha = 2.45 - 2.4 \cdot \Gamma_e, \quad 0 \leqslant \alpha \leqslant 0.85. \quad (3)$$

For models having $X < 0.3$ at the surface, representative of Wolf–Rayet stars, we adopt the mass-loss prescription by Sander et al. (2019) with the metallicity dependence proposed by Costa et al. (2021), which is based on WN and WC star models at varying Fe, C, and O abundances computed by Vink (2015). Finally, for stars in the RSG phase ($T_{eff} < 10{,}000$ K), we take the maximum between the mass-loss rates predicted by Vink et al. (2011) and de Jager et al. (1988).

### 2.2. Pulsation-driven Mass Loss

We implemented in the PARSEC code the new pulsation-driven mass loss using the results of Nakauchi et al. (2020). We denote the corresponding rate as $\dot{M}_{pdw}$. The authors performed a pulsational analysis of very massive stars with initial mass $300 \leqslant M/M_\odot \leqslant 3000$ and metallicity between $Z = 0$ and 0.002. They found their models to be unstable to radial pulsations during the early phases of the MS, as well as, when they moved to the cooler part of the HRD, during cHeB. Assuming that all pulsational energy is transferred to the mass outflow, they derived an analytic prescription for pulsation-driven mass loss. Their recipe is a function of the initial metallicity of the star and its current mass and effective temperature. To compute the mass-loss rate, they proposed two different formulations, depending on whether the star is in the MS or RSG phase during cHeB. The two equations read as follows (Equations (17) and (18) in Nakauchi et al. 2020):

$$\log\left(\frac{\dot{M}_{pdw}}{M_\odot \, \mathrm{yr}^{-1}}\right) = \alpha_1 \log\left(\frac{M}{10^3 \, M_\odot}\right) - \alpha_2 \\ - \beta_1[\log(T_{eff}) - \beta_2]^\gamma, \quad \mathrm{MS \ phase}, \quad (4)$$

$$\log\left(\frac{\dot{M}_{pdw}}{M_\odot \, \mathrm{yr}^{-1}}\right) = -2.88 + \log\left(\frac{M}{10^3 \, M_\odot}\right) \\ - 15.6[\log(T_{eff}) - 3.7], \quad \mathrm{RSG \ phase}, \quad (5)$$

where $\alpha_1$, $\alpha_2$, $\beta_1$, $\beta_2$, and $\gamma$ are coefficients that depend on the initial metallicity. Equation (4) is valid for $T_{eff} > T_{eff,min}$, a threshold also defined by the initial metallicity. Equation (5) applies for $\log(T_{eff}) \leqslant 3.85$ (3.7) when $Z \lesssim 0.0002$ (0.002). All details can be found in Table 1 of Nakauchi et al. (2020). While Equation (5) refers, strictly speaking, only to the cHeB phase, it is reasonable to assume that it can also be used to describe later phases. In fact, it is well established that cool stars tend to become increasingly unstable to pulsation as their temperature decreases (Catelan & Smith 2015). Therefore, for simplicity, we adopt Equation (5) for all post-MS stages during which the $T_{eff}$ drops below the aforementioned threshold. Because of the short duration of such phases, this assumption is expected to have a negligible impact on our results.

The results of Nakauchi et al. (2020) imply that, during the MS, the mass-loss rate increases with metallicity at a given mass. We account for this by interpolating Equation (4) in $\log(Z)$.

The stability analysis of Nakauchi et al. (2020) is strictly valid for $300 \leqslant M_i/M_\odot \leqslant 3000$, whereas we applied their mass-loss prescriptions down to $M_i = 100 \, M_\odot$. In order to validate our extrapolation, we examined the results of a few studies on pulsational mass loss in the $100 \lesssim M_i/M_\odot \lesssim 500$ range.

In order to describe the transfer of energy from pulsation to mass loss, Nakauchi et al. (2020) used the same approach as Baraffe et al. (2001), who investigated the stability of metal-free zero-age main-sequence (ZAMS) models with $120 \leqslant M/M_\odot \leqslant 500$. We find that the mass-loss rates predicted by the former authors are well compatible with the results presented in Table 3 of the latter authors for $M \geqslant 300 \, M_\odot$. At smaller masses, Equation (17) of Nakauchi et al. (2020) tends to overestimate the mass-loss rate when compared with Baraffe et al. (2001), but predictions from both studies are of the same order of magnitude. This is not expected to change the fate of the lower-mass tracks in our sets, but our $M_i = 100\, M_\odot$ model may result in a slightly larger final mass.

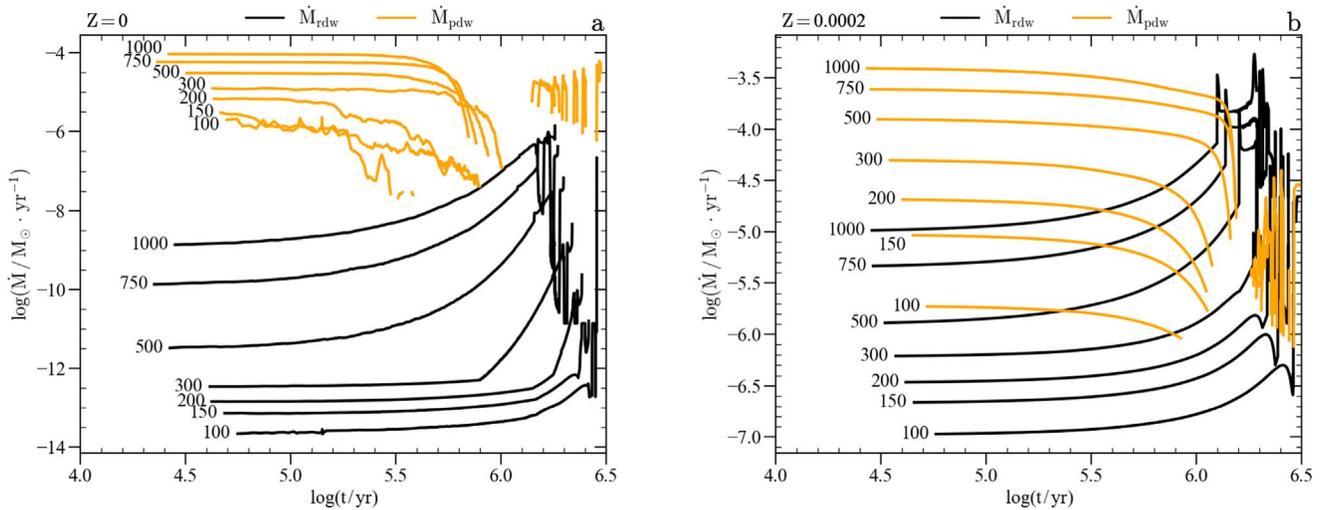

**Figure 1.** Mass-loss rate as a function of time from the ZAMS phase for tracks computed with only radiation-driven winds ($\dot{M}_{\rm rdw}$; black) and only pulsation-driven winds ($\dot{M}_{\rm pdw}$; orange). The value of the initial mass (in $M_\odot$) is indicated close to each track. Note that the vertical axis range is not the same in the two panels.

For the post-MS regime, we compared with the work by Moriya & Langer (2015), who explored the $150 \leqslant M_i/M_\odot \leqslant 250$ mass interval. Their work is also inspired by the method of Baraffe et al. (2001) but focuses on pulsation-induced mass loss after the MS. They provide analytic expressions for the mass-loss rates as a function of effective temperature and of the efficiency $\varepsilon$ with which pulsational energy is converted into kinetic energy of the outflowing matter. We considered the results for their largest conversion efficiency ($\varepsilon = 0.8$), as Nakauchi et al. (2020) effectively assumed $\varepsilon = 1$, and compared the two studies in the regime explored by Moriya & Langer (2015), that is, for $T_{\rm eff}$ approximately between 4600 and 5000 K. Over that interval of temperature, Nakauchi et al. (2020) found pulsational instability as well, and mass-loss rates predicted by both works are comparable. In particular, the mass-loss rates of Nakauchi et al. (2020) are a factor of 3–5 larger than the predictions of Moriya & Langer (2015) at $T_{\rm eff} \simeq 5000$ K, but they are smaller by nearly the same factor at $T_{\rm eff} \simeq 4600$ K, suggesting that the cumulative mass loss is of the same order of magnitude.

We note that Yadav et al. (2018) confirmed the instability found by Moriya & Langer (2015) but identified an additional regime of pulsational mass loss at higher temperatures ($\log(T_{\rm eff}) \simeq 4.2$–4.4) that they attributed to strange mode instability. The latter leads to mass-loss rates of order $10^{-7}$–$10^{-4}\, M_\odot\, {\rm yr}^{-1}$, increasing with mass, that are not predicted by Equation (18) of Nakauchi et al. (2020). It is therefore possible that our smaller-mass evolutionary tracks neglect the occurrence of mass loss during early cHeB stages, while little can be said concerning the higher masses, as the study of Yadav et al. (2018) is limited to $M \leqslant 250\, M_\odot$. However, it is reasonable to expect that strange mode instability would not cause a cumulative mass loss so large to impact our results, as it would affect relatively rapid evolutionary stages.

### 2.3. Combined Winds

Figure 1 compares the mass-loss rates associated with radiation-driven winds ($\dot{M}_{\rm rdw}$; black lines) and pulsation-driven winds ($\dot{M}_{\rm pdw}$; orange lines). The pulsation-driven mass loss dominates during the first phases of core hydrogen burning, while the radiation-driven mass loss is higher during cHeB for those tracks computed with initial metallicity $Z = 0.0002$. Given that the two types of mass loss generally dominate at different stages of evolution, we computed a set of evolutionary tracks based on the maximum mass-loss rate between $\dot{M}_{\rm rdw}$ and $\dot{M}_{\rm pdw}$, which we refer to as $\dot{M}_{\rm max}$.

In both panels of Figure 1, $\dot{M}_{\rm pdw}$ (orange lines) presents a gap at $\log(t/{\rm yr}) \sim 6$. This is because the models are evolving to lower effective temperatures during the MS phase, and at $\log(t/{\rm yr}) \sim 6$, they have $T_{\rm eff} < T_{\rm eff, min}$. As a result, Equation (4) cannot be used further, and the models temporarily stop losing mass via the pulsation-driven mechanism. When the stellar tracks cool enough and their effective temperature attains $\log(T_{\rm eff}) \leqslant 3.85$, pulsation-driven winds resume (see Equation (5)). We also notice that the models computed with $Z = 0$ exhibit an irregular behavior compared to those with $Z = 0.0002$. The cause should be linked to the scatter in the effective temperature trend, which most likely reflects some numerical noise in the convergence of the atmosphere.

Comparing Figure 1 with Figure 5 of Nakauchi et al. (2020), we note that the mass-loss rate for tracks computed with $Z = 0.0002$ is very similar during the MS phase, while in the RSG phase, the $\log(\dot{M}_{\rm pdw})$ of our stellar models is $\sim 1$ dex lower. This could be explained by the difference in effective temperature between stellar tracks in the two studies (see Section 3.1). In fact, when compared to our PARSEC tracks, the Nakauchi et al. (2020) models stretch to lower effective temperatures in the RSG phases, resulting in higher mass-loss rates (Equation (5)).

### 2.4. End of Evolutionary Calculations due to Pair Creation Instability

During the most advanced stages of massive star evolution, the electron–positron creation process absorbs some of the plasma's thermal energy, lowering the thermal pressure. As a consequence, nonideal effects enter the equation of state, preventing temperature changes from causing pressure changes. The star's layers where this process occurs become dynamically unstable. For this purpose, we use the criterion first introduced by Stothers (1999), who demonstrated that the mass-weighted average of the first adiabatic exponent, $\Gamma_1 = (\partial \log P / \partial \log \rho)_{\rm ad}$, integrated over the entire star is a useful parameter for determining a star's dynamical stability. Specifically, at each time step, we evaluate

$$\langle \Gamma_1 \rangle = \frac{\int_0^M \frac{\Gamma_1 P}{\rho} dm}{\int_0^M \frac{P}{\rho} dm}, \quad (6)$$

where $M$ is the current star's mass, $P$ is the pressure, $\rho$ is the gas mass density, and $dm$ is the mass element. The star is stable if $\langle \Gamma_1 \rangle > \frac{4}{3}$; otherwise, dynamical instability occurs. Because the PARSEC code, by construction, assumes hydrostatic equilibrium, the dynamical collapse cannot be followed. Similarly to Costa et al. (2021), we interrupt the evolution as $\langle \Gamma_1 \rangle$ falls below $4/3 + 0.01$, as first suggested by Marchant et al. (2019) and Farmer et al. (2019). All stellar models start from the ZAMS phase, progress to the end of the cHeB phase, and then ignite carbon in the core. Figure 2 shows the mass-averaged first adiabatic exponent, $\langle \Gamma_1 \rangle$, as a function of central temperature. In panels (a) and (b), most models become dynamically unstable close to the end of core carbon burning, while those with $M_i = 150$ and $100\, M_\odot$ do so after the ignition of neon or oxygen burning in the core, respectively. This happens when pair creation makes $\langle \Gamma_1 \rangle$ enter the critical regime and reach or bypass the 4/3 threshold.

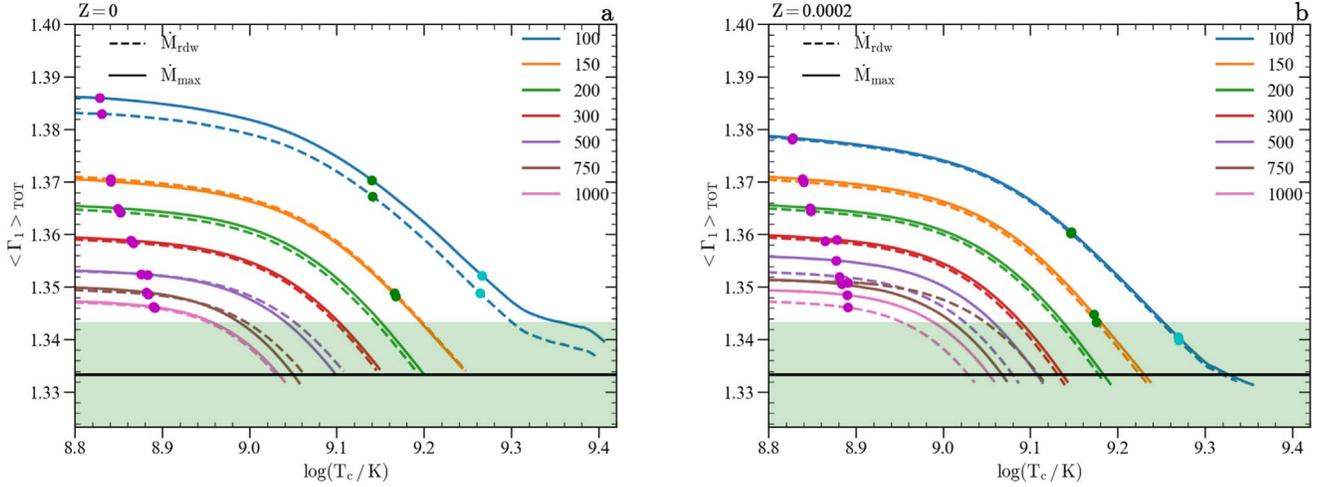

**Figure 2.** Mass-averaged first adiabatic exponent, $\langle \Gamma_1 \rangle$, as a function of central temperature and metallicity (panel (a): $Z = 0$; panel (b): $Z = 0.0002$). Results are shown for two mass-loss cases, as indicated. The magenta, green, and cyan circles mark the beginning of core C, Ne, and O burning, respectively. The thick black horizontal line denotes the threshold value of $\langle \Gamma_1 \rangle = 4/3$. The green area shows where the star can be considered dynamically unstable due to pair creation, below $\langle \Gamma_1 \rangle = 4/3 + 0.01$.

### 3. Results

#### 3.1. General Properties of the Stellar Evolutionary Tracks

Table 1 presents several quantities of the models that are key for discussing their evolution and final outcome.

Figure 3 shows all evolutionary tracks in the HRD, computed with $\dot{M}_{\rm rdw}$ and $\dot{M}_{\rm max}$. The initial metallicity influences the position of the tracks. Models with $Z = 0$ start their evolution at higher temperatures and luminosities than those with $Z = 0.0002$. The absence of metals reduces opacity and mean molecular weight, while the structures become more compact. We also recall that in massive stars at $Z = 0$, prior to the MS, due to the lack of CNO nuclei, gravitational contraction cannot be stopped until the central temperature and density are high enough for the synthesis of primary carbon via the triple-$\alpha$ process, when the star is still in the MS (e.g., Marigo et al. 2001).

Looking at the diagrams of Figure 3, we see that, during their MS, most massive tracks at $Z = 0.0002$ computed with $\dot{M}_{\rm max}$ are less luminous than those computed with $\dot{M}_{\rm rdw}$. This is most evident if we consider the bottom panels in Figure 3. At the start of the MS phase, the most massive $Z = 0.0002$ models with pulsation-driven mass loss (panel (d)) evolve almost vertically downward due to mass-loss rates as high as $10^{-4.4}$–$10^{-3.4}\, M_\odot\, {\rm yr}^{-1}$ (see Figure 1). Conversely, models with radiation-driven winds (panel (c)) suffer from very small mass-loss rates, resulting in no luminosity decrease on the MS. At $Z = 0.0002$, the luminosity difference on the MS, $\Delta \log(L) = \log(L)_{\dot{M}_{\rm rdw}} - \log(L)_{\dot{M}_{\rm max}}$, is not dramatic. It increases with $M_i$ and does not exceed 0.1 dex.

This behavior is well explained by the positive correlation between mass and luminosity on the MS with $L \propto M^\eta$, where $\eta > 0$. Compared to the standard value $\eta \simeq 3.5$–4.0 for MS stars in the range $1 \lesssim M_i/M_\odot \lesssim 10$, the mass–luminosity relation flattens out at higher masses due to the increasing contribution of radiation pressure in the central core. Following the similarity theory of stellar structure adopted by Nadyozhin & Razinkova (2005) to study the properties of very massive stars on the MS, we find that $\eta \approx 1.0$ for stars in the range $100 \lesssim M_i/M_\odot \lesssim 1000$. According to our models, the relation is a bit steeper with $1.2 \lesssim \eta \lesssim 1.6$.

After the MS, the luminosity difference, $\Delta \log(L)$, between models computed with $\dot{M}_{\rm rdw}$ and $\dot{M}_{\rm max}$ persists for tracks with $Z = 0.0002$, not exceeding $\simeq 0.15$ dex. Despite being much less pronounced, an analog luminosity difference (few 0.01 dex) affects models with $Z = 0$ as well. In general, the $\Delta \log(L)$ between each pair of tracks increases with increasing initial mass for both $Z = 0$ and 0.0002. This reflects the dependence of the pulsation-driven mass-loss rate on the current stellar mass in Equations (4) and (5).

The decrease in luminosity is also present in the stellar models of Nakauchi et al. (2020), especially in the HRD in

**Table 1**
Relevant Properties of Models Computed with $\dot{M}_{\rm rdw}$ and $\dot{M}_{\rm max}$

| $M_i$ ($M_\odot$) (1) | $\tau_{\rm MS}$ (Myr) (2) | $\tau_{\rm cHeB}$ (Myr) (3) | $f_{\rm H\,puls}$ (4) | $f_{\rm He\,puls}$ (5) | Blue Loop (6) | DUP (7) | $M_{\rm He}$ ($M_\odot$) (8) | $M_{\rm CO}$ ($M_\odot$) (9) | $M_f$ ($M_\odot$) (10) | $X_{\rm core}$ Onset PI (11) | $L_\nu/L_{\rm rad}$ ($\log_{10}$) (12) | Fate (13) | Remnant (14) | $M_{\rm BH}$ ($M_\odot$) (15) |
|---|---|---|---|---|---|---|---|---|---|---|---|---|---|---|
| | | | | | | | $Z = 0\ \dot{M}_{\rm rdw}$ | | | | | | | |
| 100 | 2.54 | 0.25 | 0.07 | 0.51 | ✓ | ✓ | 41.8 | 38.4 | 99.9 | 0.511 O | 2.7 | fCCSN[a] | BH | 89.9 |
| | | | | | | | | | | | | PPISN[b] | BH | 34.2 |
| 150 | 2.33 | 0.23 | 0.30 | 0.36 | ✓ | ✓ | 74.4 | 67.7 | 149.9 | 0.011 Ne | 3.1 | PISN | × | ⋯ |
| 200 | 2.16 | 0.22 | 0.27 | 0.35 | ✓ | ✓ | 110.4 | 103.8 | 199.9 | 0.001 C | 3.2 | PISN | × | ⋯ |
| 300 | 1.92 | 0.21 | 0.53 | 0.41 | ✓ | ✓ | 162.5 | 158.6 | 299.9 | 0.013 C | 3.2 | DBH | BH | 299.9 |
| 500 | 1.76 | 0.20 | 0.51 | 0.61 | × | ✓ | 279.2 | 270.2 | 499.9 | 0.029 C | 3.3 | DBH | BH | 499.9 |
| 750 | 1.64 | 0.19 | 0.49 | 0.64 | × | ✓ | 424.1 | 410.6 | 749.8 | 0.034 C | 3.5 | DBH | BH | 749.8 |
| 1000 | 1.59 | 0.19 | 0.54 | 0.00 | ✓ | ✓ | 565.5 | 547.0 | 999.7 | 0.028 C | 3.6 | DBH | BH | 999.7 |
| | | | | | | | $Z = 0\ \dot{M}_{\rm max}$ | | | | | | | |
| 100 | 2.59 | 0.26 | 0.17 | 0.51 | × | ✓ | 37.1 | 34.3 | 95.0 | 0.412 O | 2.8 | fCCSN[a] | BH | 85.5 |
| | | | | | | | | | | | | PPISN[b] | BH | 30.9 |
| 150 | 2.33 | 0.23 | 0.32 | 0.36 | ✓ | ✓ | 77.5 | 72.6 | 147.7 | 0.019 Ne | 3.1 | PISN | × | ⋯ |
| 200 | 2.15 | 0.22 | 0.30 | 0.35 | ✓ | ✓ | 102.9 | 95.1 | 197.6 | 0.001 C | 3.2 | PISN | × | ⋯ |
| 300 | 1.92 | 0.22 | 0.52 | 0.41 | ✓ | ✓ | 159.9 | 157.4 | 290.6 | 0.011 C | 3.3 | DBH | BH | 290.6 |
| 500 | 1.76 | 0.20 | 0.52 | 0.51 | ✓ | ✓ | 270.6 | 269.0 | 480.9 | 0.028 C | 3.4 | DBH | BH | 480.9 |
| 750 | 1.66 | 0.20 | 0.55 | 0.99 | × | ✓ | 407.6 | 380.0 | 714.5 | 0.035 C | 3.6 | DBH | BH | 714.5 |
| 1000 | 1.60 | 0.19 | 0.56 | 0.89 | ✓ | ✓ | 548.3 | 527.1 | 950.9 | 0.031 C | 3.6 | DBH | BH | 950.9 |
| | | | | | | | $Z = 0.0002\ \dot{M}_{\rm rdw}$ | | | | | | | |
| 100 | 2.83 | 0.25 | 0.29 | 0.57 | × | ✓ | 53.8 | 47.6 | 94.3 | 0.865 O | 3.1 | PPISN | BH | 40.9 |
| 150 | 2.45 | 0.24 | 0.45 | 0.32 | ✓ | ✓ | 79.5 | 71.8 | 146.4 | 0.080 Ne | 3.2 | PISN | × | ⋯ |
| 200 | 2.25 | 0.23 | 0.50 | 0.33 | ✓ | ✓ | 110.3 | 100.7 | 193.9 | 0.003 C | 3.2 | PISN | × | ⋯ |
| 300 | 2.05 | 0.22 | 0.56 | 0.77 | ✓ | ✓ | 165.8 | 150.4 | 274.2 | 0.027 C | 3.4 | DBH | BH | 274.2 |
| 500 | 1.88 | 0.20 | 0.61 | 0.44 | ✓ | ✓ | 289.1 | 265.8 | 448.9 | 0.041 C | 3.4 | DBH | BH | 448.9 |
| 750 | 1.77 | 0.21 | 0.62 | 0.99 | × | ✓ | 330.1 | 330.0 | 662.6 | 0.020 C | 2.8 | DBH | BH | 662.6 |
| 1000 | 1.71 | 0.19 | 0.60 | 0.54 | ✓ | ✓ | 575.6 | 534.6 | 831.7 | 0.029 C | 3.6 | DBH | BH | 831.7 |
| | | | | | | | $Z = 0.0002\ \dot{M}_{\rm max}$ | | | | | | | |
| 100 | 2.84 | 0.26 | 0.28 | 0.95 | × | ✓ | 53.1 | 46.9 | 92.7 | 0.859 O | 3.1 | PPISN | BH | 40.4 |
| 150 | 2.47 | 0.24 | 0.44 | 0.38 | ✓ | ✓ | 77.0 | 69.3 | 139.7 | 0.055 Ne | 3.2 | PISN | × | - |
| 200 | 2.29 | 0.23 | 0.49 | 0.36 | ✓ | ✓ | 105.7 | 96.2 | 180.2 | 0.002 C | 3.2 | PISN | × | ⋯ |
| 300 | 2.09 | 0.22 | 0.59 | 0.53 | ✓ | ✓ | 157.1 | 142.2 | 249.3 | 0.023 C | 3.3 | DBH | BH | 249.3 |
| 500 | 1.92 | 0.22 | 0.74 | 0.98 | × | ✓ | 220.6 | 207.8 | 355.8 | 0.035 C | 3.2 | DBH | BH | 355.8 |
| 750 | 1.83 | 0.21 | 0.83 | 0.39 | × | ✓ | 342.8 | 318.5 | 472.9 | 0.036 C | 3.4 | DBH | BH | 472.8 |
| 1000 | 1.77 | 0.20 | 0.85 | 0.11 | × | ✓ | 428.6 | 404.9 | 610.1 | 0.032 C | 3.4 | DBH | BH | 610.1 |

**Notes.** The columns are as follows: (1) the star's initial mass, (2) MS lifetime, (3) cHeB lifetime, (4) and (5) fractions of MS and cHeB lifetimes in which the star is unstable to radial pulsation, (6) and (7) occurrence of blue loop and DUP episode, (8) final He core mass, (9) final C–O core mass, (10) final mass of the star at the onset of dynamical instability, (11) central fuel abundance of ongoing nuclear burning at the onset of dynamical instability, (12) neutrino luminosity to radiative luminosity ratio when $T_c = 10^9$ K, (13) and (14) final fate and associated outcome (BH or complete disruption), and (15) BH mass.
[a] Failed CCSN. Following Farmer et al. (2019), we set the lower limit of $M_{\rm He}$ for PPISNe at 45 $M_\odot$.
[b] Following Woosley (2017), we set the lower limit of $M_{\rm He}$ for PPISNe at 34 $M_\odot$.

their Figure 6(c), which shows a set of tracks computed with $Z = 0.0002$. We note that the location of the ZAMS is similar in the two studies, while during the RSG phases, Nakauchi et al.'s (2020) tracks achieve lower effective temperatures than our PARSEC models.

As stars exhaust hydrogen in their cores, they evolve toward lower effective temperatures, moving to the right in the HRD. Then, during the helium-burning phase, 18 tracks out of 28 experience a blue loop; two stars evolve toward higher effective temperatures, becoming blue supergiants; and the remaining eight tracks stay at $\log(T_{\rm eff}) \sim 3.8$ as RSGs until the end of their evolution. After central He exhaustion, the stellar core contracts until it reaches the temperature required to ignite carbon. Regardless of whether the stars become dynamically unstable, the evolution after cHeB is greatly accelerated by neutrino emission (see $L_\nu/L_{\rm rad}$ in Table 1), so that the position of the tracks in the HRD does not change significantly at later stages.

When a very massive star evolves toward decreasing $T_{\rm eff}$ and approaches its Hayashi line, becoming an RSG, a DUP episode is likely to occur. While the convective envelope inflates and cools, the opacity, which is regulated by a Kramers-like law, increases so that the radiative temperature gradient exceeds the adiabatic one in the progressively deeper layers of the envelope. As a result, the bottom of the convective envelope stretches inward, passing over the H–He discontinuity, and penetrates the He core. As a consequence, helium and nitrogen, newly synthesized by the CNO cycle, are dredged up to the surface, leading to a net increase of the effective metallicity. The DUP may also occur during cHeB; in this case, the envelope may extend deeper into the developing C–O core, enriching the surface with He, C, and O. Section 3.3 examines the impact of DUP on surface abundances and chemical ejecta.

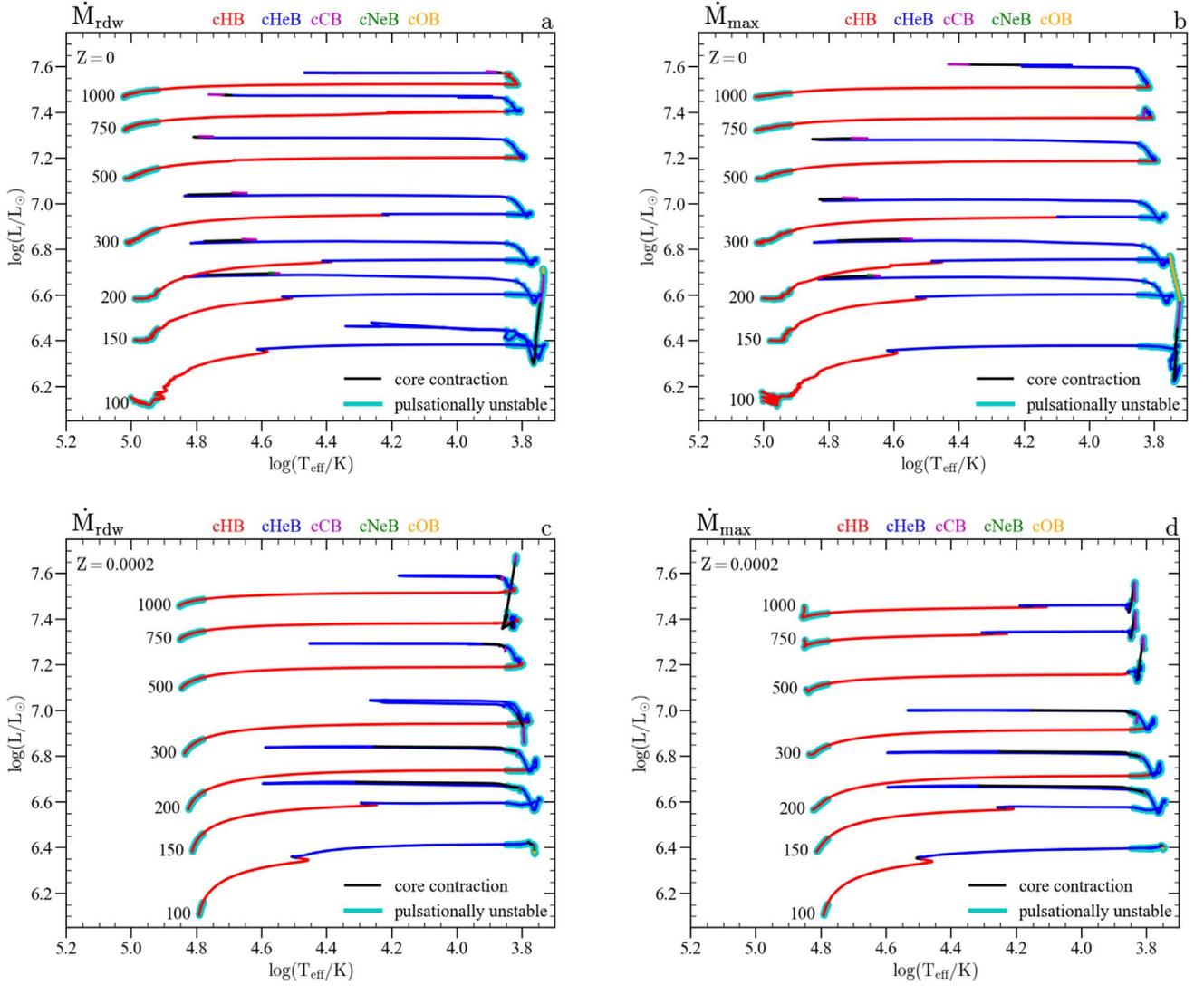

**Figure 3.** The HRDs of the four sets of tracks computed in this work. Different evolutionary phases are color-coded, as written in the legend. Panels (a) and (b) and panels (c) and (d) refer to $Z = 0$ and 0.0002, respectively. Panels (a) and (c): tracks computed with the standard PARSEC mass-loss prescription for radiation-driven winds. Panels (b) and (d): tracks computed also taking into account the pulsation-driven winds. Close to each track, the value of the initial mass (in $M_\odot$) is indicated. The cyan band superimposed on the tracks indicates where stars should be unstable against radial pulsation following Nakauchi et al. (2020).

### 3.2. Physical Overview

Here we will discuss some relevant properties of the models, with a particular focus on the physical structure.

*Evolution of the stellar center*—Figure 4 shows the central density versus central temperature diagram of all models computed with $\dot{M}_{\max}$. The general behavior of the tracks can be explained by considering the simple scaling relation

$$T_c \propto M^k \rho_c^{1/3}, \quad (7)$$

which describes the evolution of the center during a homologous contraction. The strict validity of the relation requires the fulfillment of various conditions (e.g., constant polytropic index, constant ideal gas pressure fraction, negligible thermal neutrino losses), which are usually not met by massive stars in advanced evolutionary stages. Nonetheless, the same relation may be useful to capture some fundamental dependence of the star's center evolution as a first approximation.

The exponent $k$ depends on the equation of state. If the classical ideal gas contribution dominates the total pressure, $k = 2/3$; if instead, the radiation pressure dominates the total pressure, $k = 1/3$ (Eddington 1926). We know that for a polytropic star, the ratio of the gas pressure to the total pressure, $\beta = P_{\rm gas}/P_{\rm tot}$, depends on the mass of the star,

$$\beta^{1/3}(1 + \beta) \propto M^{2/3}. \quad (8)$$

Following Zel'dovich et al. (1981) and Eddington (1926), the stellar mass at which $\beta \simeq 1/2$ roughly corresponds to $M \simeq 50\,M_\odot$. The role of radiation pressure increases with mass, so we can reasonably take $k = 1/3$ for our very massive stars. Looking at Figure 4, we can see that the tracks run almost parallel to the homologous contraction sequence (with slope 1/3), except for the last advanced stage, where some bending toward lower $T_c$ occurs, primarily driven by neutrino cooling. The factor $M^{1/3}$ in Equation (7) well explains why less massive stars reach lower $T_c$ for given $\rho_c$.

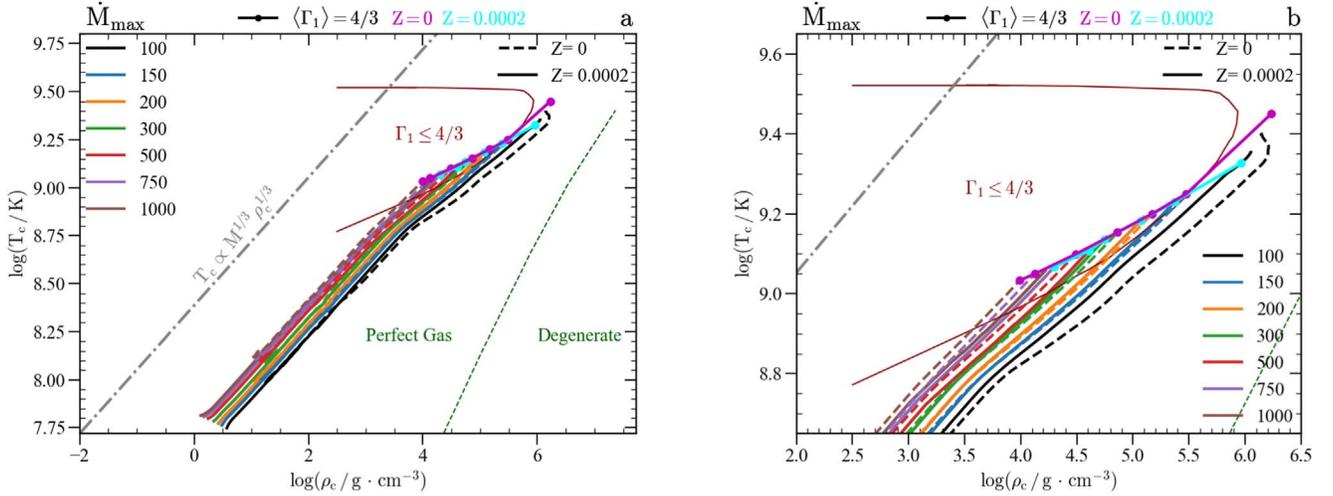

**Figure 4.** Panel (a): evolutionary tracks in the central temperature vs. central density diagram from the ZAMS to the onset of dynamical pair creation instability. Panel (b): zoom-in of the unstable region. The red curve, taken from Kozyreva et al. (2014), approximates the locus of points where $\Gamma_1 = 4/3$. The cyan (magenta) dots correspond to the points in the PARSEC $Z = 0.0002$ ($Z = 0$) models where $\langle\Gamma_1\rangle = 4/3$. The dotted–dashed line, with slope 1/3, shows the evolution during a homologous contraction.

As previously discussed in Section 2.4, most models undergo dynamical instability as a result of pair creation during carbon burning, with the exception of the $M_i = 100$ and $150\,M_\odot$ models, which experience this condition later, after the onset of oxygen and neon burning, respectively. This is evident in the $\rho_c$–$T_c$ diagram, as almost all of the tracks enter the $\Gamma_1 < 4/3$ region (red curve; taken from Kozyreva et al. 2014). The models with $M_i = 100\,M_\odot$ do not appear to cross the critical boundary, whereas the integration of Equation (6) yields the opposite result (cyan and magenta circles). The apparent discrepancy is misleading. In fact, the red curve in Figure 4 defines the locus where $\Gamma_1 = 4/3$ in the center, whereas our $M_i = 100\,M_\odot$ models experience off-center pair creation and enter the unstable region, as illustrated in Figure 2.

Figure 5 shows the Kippenhahn diagrams of a few selected models computed with $\dot{M}_{\rm rdw}$ (left panels) and $\dot{M}_{\rm max}$ (right panels). One distinguishing feature of very massive stars is that, even in the absence of rotation or other mixing processes, they evolve nearly homogeneously during the MS phase because they develop very large convective cores, initially covering up to $\approx$80% of the total mass. As hydrogen is burned, convective cores gradually recede due to the significant contribution of the radiation pressure $P_{\rm rad}$ ($P/T^4 \propto P/P_{\rm rad} \propto (1-\beta)^{-1}$) and decreasing electron scattering opacity ($\kappa_{\rm es} \simeq 0.2\,(1+X)\,{\rm cm}^2\,{\rm g}^{-1}$). Both factors concur to lower the radiative temperature gradient. The MS lifetime ranges from $\simeq$2.8 to 1.6 Myr passing from $M_i = 100$ to $1000\,M_\odot$.

The fraction of the MS lifetime where pulsation instability occurs is significant, as shown in Table 1, and it increases with stellar mass and metallicity. For example, the ($M_i = 1000\,M_\odot$, $Z = 0.0002$, $\dot{M}_{\rm max}$) model experiences radial pulsation and associated mass loss for $\approx$85% of its MS phase. As a result, the reduction in stellar mass ($\dot{M}_{\rm max}$ case) is much greater than in the case of weak radiative winds ($\dot{M}_{\rm rdw}$ case). For the former model, the stellar mass at the end of the MS phase is $M = 611\,M_\odot$, while the latter has $M = 885\,M_\odot$. During the subsequent cHeB phase, all tracks develop convective cores. The cHeB lifetimes are roughly 10% the MS duration, as expected.

In Section 3, we mentioned the possibility of a star experiencing DUP episodes as it approaches its Hayashi line. This is most common during the cHeB phase (see the H-R tracks of Figure 3), when the envelope extends deeper into the He or C–O core. This occurs, for example, in the $M_i = 100\,M_\odot$ models at $Z = 0$ (panels (a) and (b)). Using $\dot{M}_{\rm rdw}$, the star experiences a first DUP that enters the He core, followed by a second DUP that extends into the forming C–O core. As we will see in Section 3.3, the first mixing episode causes a dramatic increase in N at the surface, whereas the second episode enriches the envelope primarily with C and O. The same phenomenon occurs in the $M_i = 100\,M_\odot$ with $\dot{M}_{\rm max}$, but in this case, the envelope deepens more gradually. Similar considerations apply when comparing the ($M_i = 750$, $1000\,M_\odot$; $Z = 0.0002$) models, computed with $\dot{M}_{\rm rdw}$ and $\dot{M}_{\rm max}$.

All structures displayed in Figure 5 become dynamically unstable due to the pair creation in different stages of evolution (see Section 2.4 and Table 1).

### 3.3. Surface Chemical Abundances

Here we discuss the evolution of the surface abundances that can be modified by DUP episodes, as well as the composition of the chemical ejecta, which is also affected by stellar wind efficiency.

Figure 6 shows the surface abundance evolution of a few relevant nuclides in some selected models. Each panel compares the results for a model with the same initial mass obtained with two mass-loss prescriptions, namely, $\dot{M}_{\rm rdw}$ (dotted lines) and $\dot{M}_{\rm max}$ (solid lines).

In Section 3.1, we have already discussed the main characteristics of these mixing events, especially in relation to the Kippenhahn diagrams (Figure 5), to which the reader should refer for better comprehension. All models depicted in Figure 6 experience a DUP episode but with varying degrees of envelope penetration. Very deep DUPs occur in the ($M_i = 100\,M_\odot$, $Z = 0$) and ($M_i = 750\,M_\odot$, $Z = 0.0002$, $\dot{M}_{\rm rdw}$) models, as discussed below.

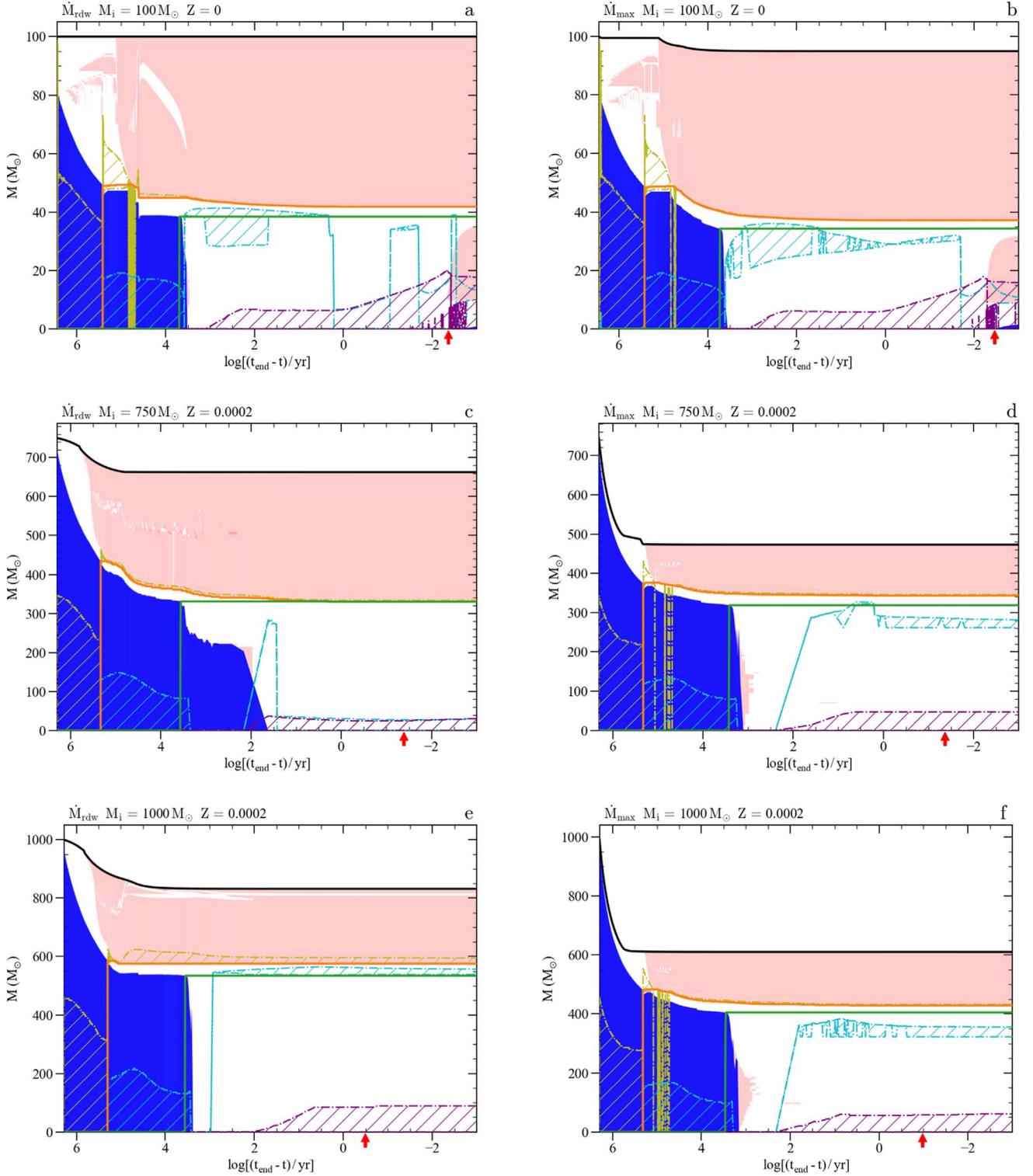

**Figure 5.** Kippenhahn diagrams of selected models. The horizontal axis represents the logarithm of time (in years) until the onset of pair creation dynamical instability. The blue regions in each diagram represent the star's convective core, and the pink areas correspond to the convective envelope, semiconvective zones at the boundary of the helium convective core, and convective shells. The yellow, cyan, and purple hatched regions represent the hydrogen-, helium-, and carbon-burning core/shells, respectively. The black solid line shows the total mass of the star, the orange line corresponds to the helium core, and the green line indicates the carbon–oxygen core. The red arrow marks the time when the star enters the unstable region with $\langle \Gamma_1 \rangle = 4/3 + 0.01$. Panels (a), (c), and (e): models computed with the standard PARSEC mass-loss prescription for radiation-driven winds. Panels (b), (d), and (f): models that include pulsation-driven winds.

In general, the models with DUP display a surface depletion of H and an increase in $^4$He, $^{14}$N, $^{12}$C, and $^{16}$O. When the envelope crosses the H–He discontinuity and enters the He core, which contains the products of complete H burning through the CNO cycle, $^4$He and $^{14}$N are enriched at the surface. This situation is best illustrated (panel (a)) by the ($M_i = 100\,M_\odot$, $Z = 0$, $\dot{M}_{\rm rdw}$) model, where there is a sudden and significant increase in $^{14}$N just as the bottom of the convective envelope stretches into the He core. At the same time, despite the fact that the CNO cycle depletes $^{12}$C and $^{16}$O in the He core in favor of $^{14}$N, their surface abundances increase because the material extracted from the He core is diluted in the envelope, which initially contains no metals ($Z = 0$).

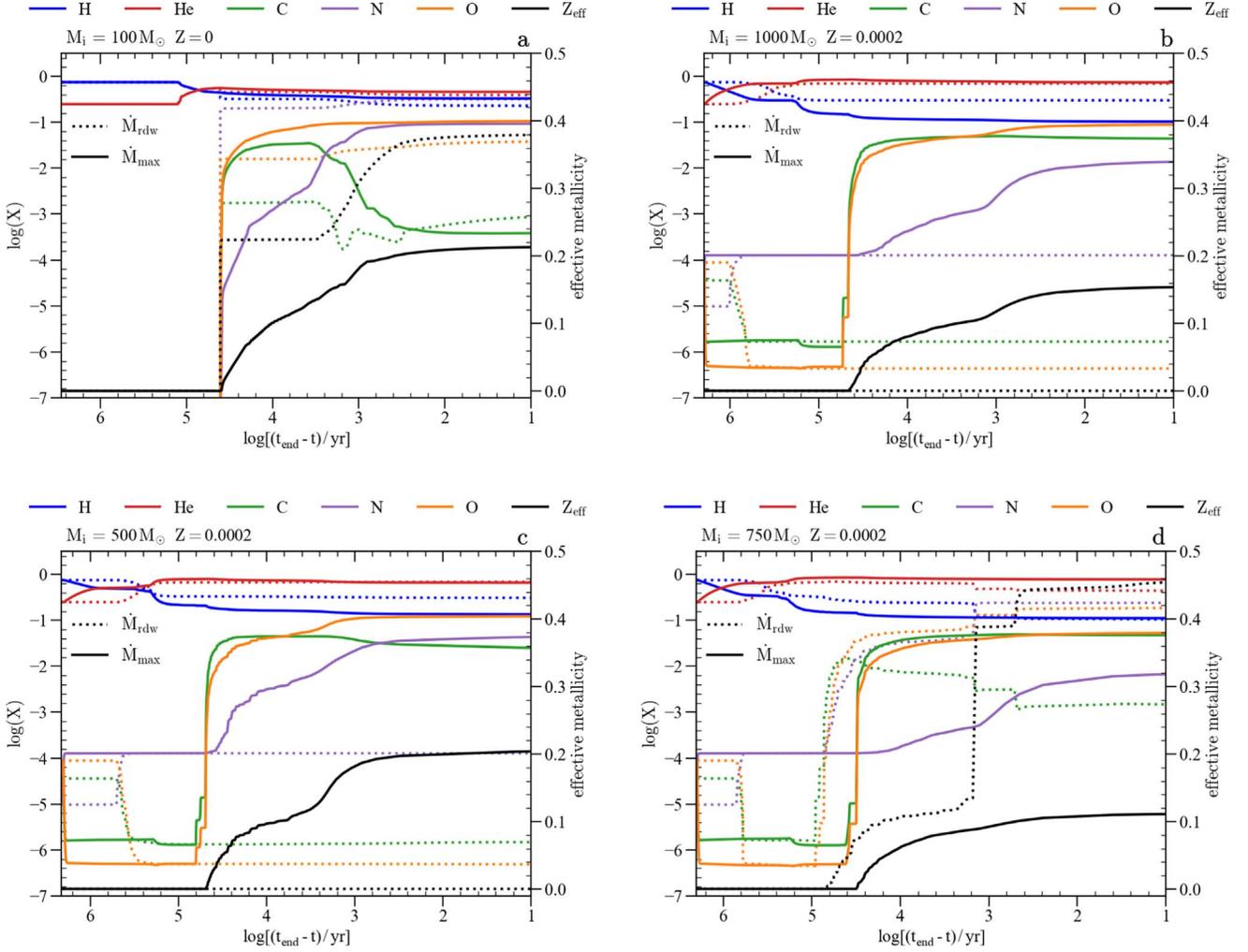

**Figure 6.** Evolution of surface chemical abundances of four selected models from the ZAMS to the onset of dynamical instability. In each panel, the abundances of five elements, namely, hydrogen, helium, carbon, nitrogen, and oxygen, are depicted in different colors. The effective metallicity, $Z_{\rm eff}$, is shown in black. The results are presented for two different mass-loss prescriptions. The horizontal axis is the logarithm of time (in years) until the onset of dynamical instability.

As the envelope deepens, its base may even reach and enter the forming C–O core, e.g., the ($M_{\rm i} = 750\,M_\odot$, $Z = 0.0002$, $\dot{M}_{\rm rdw}$) model in Figure 5(c), where $^{4}$He is burned into $^{12}$C and $^{16}$O, while $^{14}$N is gradually converted into $^{22}$Ne via the chain $^{14}{\rm N}(\alpha,\gamma)^{18}{\rm F}(\beta^+\nu)^{18}{\rm O}(\alpha,\gamma)^{22}{\rm Ne}$. As we can see, the chemical enrichment in the ($M_{\rm i} = 100\,M_\odot$, $Z = 0$, $\dot{M}_{\rm max}$) model is more gradual than the analog for $\dot{M}_{\rm rdw}$, and it misses the abrupt initial jump in $^{14}$N abundance. Overall, differences in surface abundance evolution between models with the same initial mass but different mass-loss rates reflect differences in chemical profiles, opacity, and convective border details. The DUP results in a net increase in surface effective metallicity ($Z_{\rm eff} = 1 - X - Y$; dotted/solid black lines). The case of the ($M_{\rm i} = 100\,M_\odot$, $Z = 0$, $\dot{M}_{\rm rdw}$) model is particularly noteworthy, with $Z_{\rm eff}$ increasing as high as 0.38 owing to the large $^{14}$N abundance. We note that for both mass-loss prescriptions, the occurrence of the DUP significantly reduces the He core mass, which passes from $M_{\rm He} \simeq 48\,M_\odot$ at the end of the MS to $M_{\rm He} \simeq 37\,M_\odot$ with $\dot{M}_{\rm max}$ and $M_{\rm He} \simeq 42\,M_\odot$ with $\dot{M}_{\rm rdw}$. Such reduction is especially important for the final outcome of these models (see Section 3.4).

Figure 7 presents the chemical ejecta of He, C, N, O, Ne, and Mg for a few models. Each panel contains the results of two mass-loss prescriptions, as indicated in the legend. Tables of wind ejecta can be found on Zenodo: doi:10.5281/zenodo.7528650. Chemical ejecta computed with $\dot{M}_{\rm rdw}$ are, as expected, lower than those computed with $\dot{M}_{\rm max}$, since the latter takes the maximum of ($\dot{M}_{\rm rdw}$, $\dot{M}_{\rm pdw}$) by construction. We verified that the main difference in the ejected mass of all of the considered elements is caused by mass loss for each pair of tracks in Figure 7. The only exception is nitrogen, whose ejecta is higher in the ($M_{\rm i} = 100\,M_\odot$, $Z = 0$) model with $\dot{M}_{\rm rdw}$ due to a deeper DUP episode. Furthermore, while wind ejecta are greater for $Z = 0.0002$ than for $Z = 0$, we can see that, regardless of metallicity, large amounts of helium, up to several tens of or a few hundred solar masses, are expelled from the most massive models.

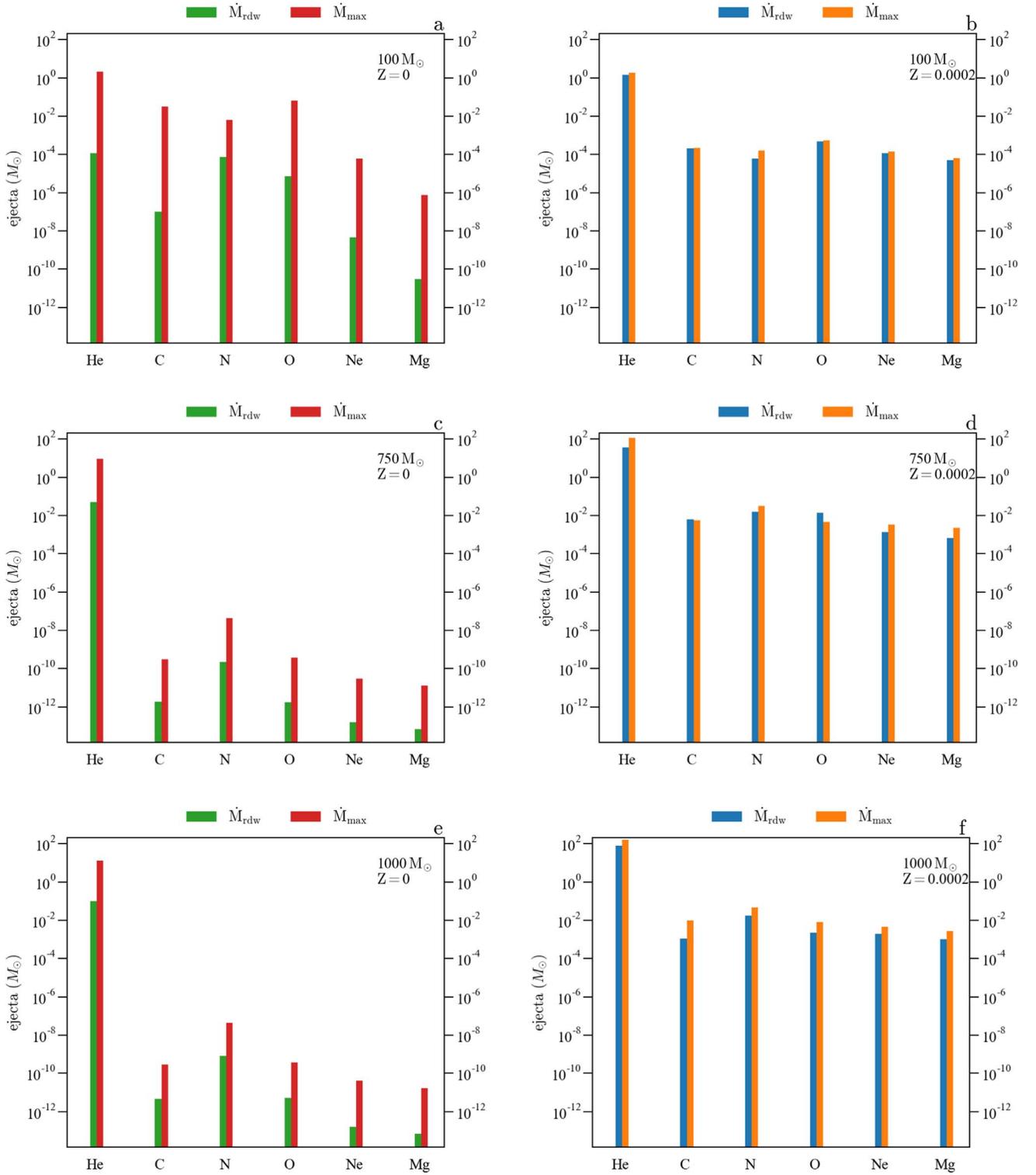

**Figure 7.** Ejecta mass of models with three selected initial mass, namely, $M_i/M_\odot = 100$, 750, and 1000. Each panel shows the ejecta mass of helium, carbon, nitrogen, oxygen, neon, and magnesium for two mass-loss recipes, $\dot{M}_{\rm rdw}$ and $\dot{M}_{\rm max}$. Panels (a), (c), and (e): models computed with $Z = 0$. Panels (b), (d), and (f): models computed with $Z = 0.0002$.

### 3.4. Final Fate

Figure 8 (left panel) shows the helium core mass, $M_{\rm He}$, at the onset of pair creation dynamical instability for all tracks. Overall, there is a positive correlation between $M_i$ and $M_{\rm He}$. A deep DUP causes a sudden change in slope at $M_i = 750\,M_\odot$ in the ($Z = 0.0002$, $\dot{M}_{\rm rdw}$) sequence. When we consider tracks of the same metallicity, we can see that models computed with $\dot{M}_{max}$ end up with smaller $M_{He}$. The effect is more pronounced at $Z = 0.0002$ because the mass reduction on the MS through $\dot{M}_{max}$ is much stronger than with $\dot{M}_{rdw}$.

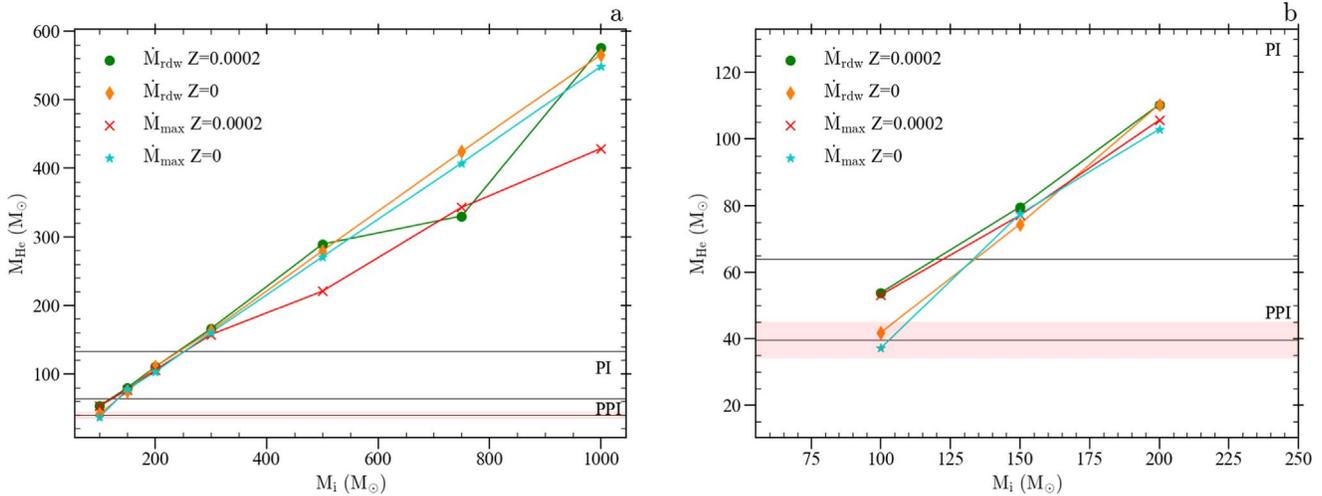

**Figure 8.** Panel (a): helium core mass, $M_{He}$, as a function of the initial mass for all models of four different sets, as stated in the legend. Helium core mass is evaluated at the onset of dynamical instability. Horizontal lines delimit the regimes in which pulsation pair instability, pair instability explosion, and direct collapse to BHs are expected (Woosley 2017; Farmer et al. 2019, 2020). Panel (b): zoom-in of the $100 \lesssim M_i/M_\odot \lesssim 200$ range. The red strip indicates the uncertainty range of the lower limit for pulsation pair instability. Lower and upper boundaries are 34 (Woosley 2017) and 45 (Farmer et al. 2019) $M_\odot$, respectively. The black line is the average.

We can see that, regardless of metallicity or mass-loss prescription, stars with $M_i \gtrsim 300 M_\odot$ should avoid the PISN channel and collapse directly to a BH. In fact, their He cores have masses that exceed the limit of $130 \lesssim M_{He}/M_\odot \lesssim 133$–$139$ (Woosley et al. 2007; Woosley 2017; Farmer et al. 2020; Woosley & Heger 2021). Instead, stars with $M_i = 150$ and $200 M_\odot$ have He cores that are just massive enough to cause the pair instability explosion, leading to total disruption. Finally, stars with $M_i = 100 M_\odot$ should experience pulsation pair instabilities or end as failed CCSNe, resulting in the formation of a BH as a compact remnant.

To a first approximation, the final mass, $M_f$, at the end of the hydrostatic evolution provides a rough estimate of the remnant BH mass for stars that undergo DBH. While the BH mass for a PISN is simply zero, for PPISNe, we use the formula proposed by Spera & Mapelli (2017; with the corrections of Mapelli et al. 2020), which fits the results of Woosley's (2017) hydrodynamic calculations. We also account for the mass loss due to the neutrino emission, which we set equal to $0.1 M_{bar}$, where $M_{bar}$ is the baryonic mass of the proto–compact object (Fryer et al. 2012; Rahman et al. 2022, and references therein). The PPISN configuration applies to the ($M_i = 100 M_\odot$, $Z = 0.0002$) models, whereas for the ($M_i = 100 M_\odot$, $Z = 0$) models, the fate is somewhat uncertain. The star with $\dot{M}_{max}$ has a helium core mass of $\simeq 37 M_\odot$. We can assess its outcome by comparing $M_{He}$ to the lower limit for the development of pulsation pair instabilities. According to Woosley (2017), the threshold is around $34 M_\odot$, while Farmer et al. (2019) indicated that it is about $45 M_\odot$.

On the one hand, if we follow Woosley (2017), the ($M_i = 100 M_\odot$, $Z = 0$, $\dot{M}_{max}$) star should be able to enter the PPISN regime, producing a BH of mass $\simeq 30.9 M_\odot$. On the other hand, if we follow Farmer et al. (2019), the same star should avoid the PPISN path and complete the entire sequence of nuclear burnings up to the formation of an iron core, which eventually collapses, resulting in a failed CCSN, assuming efficient fallback (Fryer et al. 2012; delayed model). The estimated BH mass would be $\simeq 85.5 M_\odot$, under the hypothesis that $\simeq 0.1 M_f$ is lost due to neutrino emission.

We observe that the analysis of Farmer et al. (2019) relies on pure He models, while our calculations follow the evolution of complete models. In this respect, we note that the stability analysis based on $\langle \Gamma_1 \rangle$ is primarily controlled by the core mass and its chemical composition, with a small influence from the residual envelope (e.g., Costa et al. 2021, Tables A1 and A2). As a result, taking the lower threshold limit of Farmer et al. (2019) is still a reasonable assumption for our exploratory study. In general, differences in the mass limits of the pair creation instability window reflect differences in the input physics among various sets of models.

Based on the dense grid of PARSEC models computed by Costa et al. (2021), the lower limit for entering the PPISN regime is $M_{He} \simeq 36$–$39 M_\odot$ if we take the threshold for the onset of PI at $\langle \Gamma_1 \rangle = 4/3 + 0.01$. This value is roughly halfway between the boundaries indicated by Woosley (2017) and Farmer et al. (2019). If we take the threshold for the onset of PI strictly at $\langle \Gamma_1 \rangle = 4/3$, the lower boundary for PPISNe in the Costa et al. (2021) models shifts at $M_{He} \simeq 48 M_\odot$.

Our structure calculations of the ($M_i = 100 M_\odot$, $Z = 0$, $\dot{M}_{max}$) track suggest that during the onset of O burning, the mass-averaged $\langle \Gamma_1 \rangle$ is approaching the critical value of $4/3$ due to pair creation. If this threshold was exceeded at some later stage ($\langle \Gamma_1 \rangle < 4/3$), then the star would enter the PPISN regime. Similar considerations apply to the ($M_i = 100 M_\odot$, $Z = 0$, $\dot{M}_{rdw}$) model.

Table 1 and Figure 9 compare the results obtained with different mass-loss prescriptions. Our calculations show that low-metallicity very massive stars can produce BHs with masses exceeding $\sim 100 M_\odot$. The most massive BHs are produced by very massive stars with $Z = 0$, as mass loss is modest. The final mass of our models with $300 \leqslant M_i/M_\odot \leqslant 1000$ is higher with respect to those of Nakauchi et al. (2020, Figure 8). At $Z = 0$, the difference is at most $\sim 7\%$, while for models computed with $Z = 0.0002$, the difference is at most $\sim 24\%$. This reflects the difference in mass-loss rates between the two sets of models, which is mostly caused by the different evolution of the effective temperature (see Sections 2.3 and 3.1).

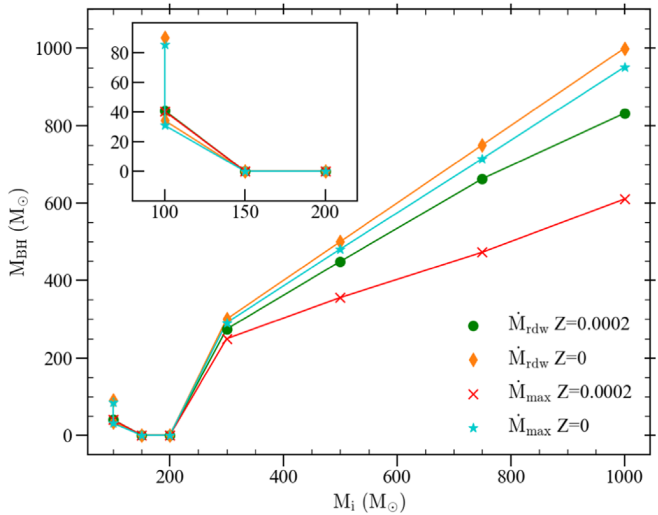

**Figure 9.** The BH mass as a function of the initial mass for all tracks presented in this work. The inset plot zooms in on the $100 \leqslant M_i/M_\odot \leqslant 200$ range. The cyan stars and orange diamonds correspond to the ($M_i = 100\,M_\odot$, $Z = 0$, $\dot{M}_{max}$) and ($M_i = 100\,M_\odot$, $Z = 0$, $\dot{M}_{rdw}$) models, respectively. They indicate the predicted BH masses, depending on whether the star results in a failed CCSN or PPISN. See also Table 1.

## 4. Concluding Remarks

In this study, we investigate the evolution of zero-metallicity ($Z = 0$) and extremely metal-poor ($Z = 0.0002$) very massive stars with initial masses ranging from 100 to 1000 $M_\odot$. These calculations extend the PARSEC evolutionary models in the very high mass regime. One novel element is the inclusion of pulsation-driven winds, following the findings of a recent study (Nakauchi et al. 2020), in which a stability analysis against radial pulsation is performed. In addition to pulsation-driven mass loss, we also consider the occurrence of radiation-driven winds in both the hot and cool regions of the HRD. We find that the two mechanisms prevail at different stages. In particular, the amount of mass ejected by the most massive models through pulsation instability during the MS can be substantial, in contrast to the modest outflows expected via momentum absorption from radiation.

All models are followed until the onset of dynamical instability caused by the creation of electron–positron pairs, which occurs near the end of core carbon burning or shortly after the start of neon or oxygen burning, depending on the initial mass. We extract the criterion to assess the final outcome of the models and the type of compact remnant based on full computations (Kozyreva et al. 2014; Woosley 2017; Farmer et al. 2019).

We find that stars with $M_i \geqslant 300\,M_\odot$ should end their lives without exploding, instead directly collapsing to BHs. Our models with $M_i = 150$ and $200\,M_\odot$ should produce PISNe, leaving no remnant and thus contributing to the primordial BH mass gap.

Depending on metallicity and mass loss, models with $M_i = 100\,M_\odot$ may have a different fate. At $Z = 0.0002$, they should enter the PPISN window, ejecting some mass before collapsing to BHs. At $Z = 0$, the outcome is somewhat uncertain. The $M_i = 100\,M_\odot$ model could die as either a failed CCSN or a PPISN, depending on the predicted width of the PPISN strip (see the discussion in Section 3.4).

In the event of a failed CCSN, the remnant BH mass of $\simeq 85.5\,M_\odot$ is very close to the estimated primary BH mass of $85^{+21}_{-14}\,M_\odot$ for the binary BH merger GW190521 (Abbott et al. 2020). We may speculate that primordial very massive stars with $M_i \simeq 100\,M_\odot$ could help us alleviate the BH mass gap conundrum (for an overview, see Costa et al. 2021). The failed CCSN associated with our ($M_i = 100\,M_\odot$, $Z = 0$) models provides another possible pathway for the formation of BHs with masses between 40–65 and 120 $M_\odot$ (see also Croon et al. 2020; Farmer et al. 2020; Sakstein et al. 2020; Costa et al. 2021; Farrell et al. 2021; Tanikawa et al. 2021; Vink et al. 2021; Farag et al. 2022, for similar conclusions).

Another key process of very massive star evolution is rotation (Yusof et al. 2013; Goswami et al. 2022; Higgins et al. 2022), which will be investigated in a follow-up work. If enough angular momentum was retained in their cores, these very massive stars could produce gamma-ray bursts, known as supercollapsars (Woosley 1993; Yoon et al. 2012).

On the observational side, the James Webb Space Telescope (JWST) will open a new window on Population III stars. Since isolated primordial stars are likely not accessible to JWST, small Population III galaxies and their integrated colors may provide the best opportunities for directly probing the properties of metal-free stars (Zackrisson et al. 2011). Furthermore, thanks to their enhanced sensitivity, future ground-based (Einstein Telescope, Cosmic Explorer) and space-based (LISA, DECIGO) detectors are expected to collect gravitational-wave events from binary BH mergers in the range of $\approx 10^2$–$10^4\,M_\odot$ up to a redshift of $\approx 20$ (Fragione et al. 2022; Saini et al. 2022), a regime so far unexplored. In this perspective, theoretical studies of the evolution of primordial very massive stars are critical for contextualizing the upcoming data within an astrophysical picture.

P.M., G.V., M.T., and L.G. acknowledge support from Padova University through the research project PRD 2021. A. B. acknowledges support by PRIN MIUR 2017 prot. 20173ML3WW 001. G.C. acknowledges financial support from the European Research Council for the ERC Consolidator grant DEMOBLACK under contract No. 770017.

*Software:* PARSEC (Bressan et al. 2012; Costa et al. 2019, 2021), OPAL (Iglesias & Rogers 1996), ÆSOPUS (Marigo & Aringer 2009), FREEEOS (A. W. Irwin; http://freeeos.sourceforge.net).

## ORCID iDs

Guglielmo Volpato ● https://orcid.org/0000-0002-8691-4940
Paola Marigo ● https://orcid.org/0000-0002-9137-0773
Guglielmo Costa ● https://orcid.org/0000-0002-6213-6988
Alessandro Bressan ● https://orcid.org/0000-0002-7922-8440
Michele Trabucchi ● https://orcid.org/0000-0002-1429-2388
Léo Girardi ● https://orcid.org/0000-0002-6301-3269